\documentstyle[amssymb,prl,aps,twocolumn,epsf]{revtex}

\newcommand{\be}{\begin{equation}}
\newcommand{\ee}{\end{equation}}
\newcommand{\bea}{\begin{eqnarray}}
\newcommand{\eea}{\end{eqnarray}}

\begin{document}

\twocolumn[\hsize\textwidth\columnwidth\hsize\csname @twocolumnfalse\endcsname

\title {Theory of Microemulsion Glasses}
\author{Sangwook Wu$^{(a)}$, H. Westfahl Jr.$^{(a)}$, J. Schmalian$^{(a)}$ and
 P. G. Wolynes$^{(b)}$}
\address{$^a$ Department of Physics and Astronomy and Ames Laboratory, Iowa  State
University, Ames, IA 50011\\
$^b$ Department of Chemistry and Biochemistry,
University of California at San Diego, La Jolla, CA 92093}
\date{\today}
\draft
\maketitle
\begin{abstract}
 We show that the tendency towards microphase separation in microemulsions
leads to the formation of a glassy state after sufficiently strong correlations between polar and hydrophobic regions
have been established. Glassiness is predicted to occur above a critical volume fraction of surfactant,
which is determined by the length of the amphiphilic molecules. Our results are obtained by solving the dynamical equations for
the correlation functions of the system and by using a replica approach.
\end{abstract}
] \narrowtext

Oil and water phase separate at low temperatures, an effect which can be
altered by adding amphiphilic surfactant molecules like soap or lipids.
Depending on the nature of the surfactant and its volume fraction, complex
inhomogeneous structures occur\cite{GT83}. These are caused by the competition between
short-ranged forces between oil and water, favoring the separation of
uniformly condensed phases, and long-range forces due to the surfactant
which energetically frustrate this separation. Examples of such structures
are emulsions, which are nonequilibrium colloidal suspensions, and microemulsions
in which oil and water are intertwined in complex structures but at equilibrium.
The former consist of macroscopically large droplets or bicontinuous networks
of oil and water separated by monolayer interfaces of amphiphiles.
Microemulsions, on the other hand, are composed of self-organized mesoscopic structures in form of micelles,
vesicles or lamellae. These strongly correlated fluids are of great
scientific and technological interest: they present extreme materials
properties, like ultra-small surface tensions; they are essential for the
stability of cell membranes, formed by phospholipid molecules; their
applications range from medicine to biomolecular assemblies such as the Golgi apparatus, to
food science in the preparation of sauces,
and to petroleum industry, just to name a few prominent examples.

Many of the mesoscale structures found in these amphiphilic systems are extremely long lived i.e.,
the dynamics is glassy. Sometimes, the macroscopic mechanical properties are like those
of a soft solid, as in ``stiff mayonnaise". Light and neutron scattering reveal the hallmarks
of glassy motions also on the mesoscopic scale \cite{Sheu89,Choy00,Gang99} of amphiphilic assemblies.
While this glassy behavior is in many ways analogous to that of vitrifying molecular liquids,
it goes on at a larger length scale which provides both new opportunities for probing glass physics, and
as we shall see for tuning it.

In this paper we develop a theory for glassiness in oil-water-surfactant
mixtures for the case of equal water and oil volume fractions, using an
electrostatic analogy\cite{Stillinger83,Deem94} to describe the competition
between entropic effects and stoichiometric constraints of amphiphiles. Although
the equal volume fraction case is easier to analyze, our study suggests that little
is changed for unbalanced compositions. Glassy behavior in those cases has been attributed
to the dynamics of random packed pre-formed micelles or inverse micelles.

\begin{figure}[t] \begin{center} \leavevmode \epsfxsize=8.5cm
\epsffile{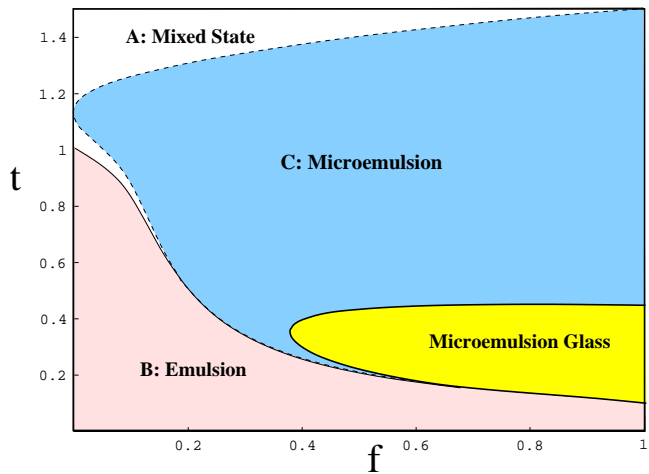} \end{center} \caption{Phase diagram of the
water-oil-surfactant mixture for $r_s=10a$, with $t$ as the reduced temperature and
$f$ the surfactant volume fraction. The dashed line represents a
crossover between the mixed (A) and the microemulsion (B) phases, while
the full line separating the emulsion phase (C) from the
other phases represents a phase transition. The microemulsion glass transition
appears on the region where the correlations between the micellar structures
are strong} \label{fig1}
\end{figure}

The main result of this paper is summarized in the phase diagram, Fig.1. We
find that a self generated microemulsion glass forms above a critical volume
fraction of the surfactant, $f$, provided that the mean square root length, $%
r_{{\rm s}}$, between the head and the tail parts of the surfactant molecule
exceeds a minimum value (at least $\ 6$ times the typical correlation length
of the water and oil molecules, $a$). The emergence of glassiness is
determined by solving the dynamical equations for the density-density
correlation and response functions of the system\cite{CK93} as well as by
using a replica approach for systems without quenched randomness\cite
{Mon95,MP991}. In the dynamical approach, glassiness is manifested in the
form of anomalous long time dynamics and aging effects. In the replica
approach, it is reflected in the emergence of an exponentially large number
of metastable states ${\cal N}_{{\rm ms}}$ which renders the system unable
to behave as a fluid, resulting in the divergence of the viscosity,
accompanied by a loss of the configurational entropy $S_{c}=-k_{{\rm B}}\log
$ ${\cal N}_{{\rm ms}}$. Both approaches yield identical criteria for the
emergence of glassiness. In addition to the glassy state, we also find, in
agreement with earlier studies\cite{Woo95}, a region of volume fractions and
temperatures where the system forms emulsions and a region where a fluid of
microemulsions, built of small micelles or lamellar structures, forms.
Finally, for large temperatures there is a region where the mixing entropy
favors a homogeneous state.

Self assembled lamellar or micellar structures form in water-oil mixtures
due to the competition between entropic effects and constraints caused by
the stoichiometry of the amphiphilic surfactants, which are hydrophobic at
one end and hydrophilic (polar) at the other end. An elegant description of
this complex competition is achieved by using an electrostatic analogy.\cite{Stillinger83,Deem94}
Stoichiometry enforces that the averaged densities of
polar and hydrophobic surfactant groups have to be the same over regions of
the size of the mean square root length of the surfactant molecule, $r_{{\rm %
s}}$. This is achieved if one introduces a fictitious Coulomb potential, see
Eq.\ref{ham11} below, where the effective charges are positive (negative)
for hydrophobic (polar) molecules. $r_{{\rm s}}$ acts as the screening
length of this potential since it is the scale on which charge neutrality
must be enforced. In Ref. \cite{Woo95} the relevant parameters of the
electrostatic model have been expressed in terms of microscopically well
motivated physical quantities like the oil-water surface tension, $\sigma $,
the correlation length of the water and oil molecules, $a$, the volume
fraction of the surfactant, $f$, as well as $r_{{\rm s}}$. In what follows
we use a version of the theory, obtained by (approximately) integrating out
the surfactant degrees of freedom, which leads to an effective interaction
between all hydrophilic and all polar molecules of the system.

A system without surfactants which consists in equal parts of oil-like and
water-like molecules is described by the energy
\begin{equation}
E_{0}=\frac{1}{2}\int d^{d}x\left( -\frac{6}{a^{2}}\rho ^{2}+\left( \nabla
\rho \right) ^{2}+\frac{u}{2}\rho ^{4}\right) ,  \label{ham00}
\end{equation}
where $\rho ({\bf x})>0$ in a region with hydrophobic molecules and $\rho (%
{\bf x})<0$ for polar molecules and $\left\langle \rho ({\bf x}%
)\right\rangle =0$ on the average. Macroscopic phase separation occurs at a
temperature $T_{c}^{0}=3\sigma a^{2}$ where the regions with opposite sign
of $\rho $ are separated by an interface with surface tension $\sigma $,
determined by $a$ as well as the quartic interaction, $u$. If water and oil
occur in unequal parts, additional odd powers of $\rho $ occur. For
simplicity, here we neglect these effects, although the generalization of
our approach is straightforward.

Adding surfactant molecules with volume fraction, $f$, frustrates this
macroscopic phase separation and results in an effective interaction between
the hydrophobic and polar species \cite{Woo95}
\begin{equation}
{\cal \ }V=\frac{9ft}{2\pi r_{{\rm s}}^{3}a}\int d^{d}x\int d^{d}x^{\prime }%
\frac{e^{-\left| {\bf x-x}^{\prime }\right| q_{{\rm D}}}\rho ({\bf x)}\rho (%
{\bf x}^{\prime })}{\left| {\bf x-x}^{\prime }\right| }.  \label{ham11}
\end{equation}
Here $t=T/T_{c}^{0}$ is the reduced temperature of the system and, as
expected, $q_{{\rm D}}=6^{1/2}r_{{\rm s}}^{-1}$ \ plays the role of a Debye
wave vector with screening length $\varpropto r_{{\rm s}}$.

\begin{figure}[t]
\begin{center}
\leavevmode \epsfxsize=8.5cm \epsffile{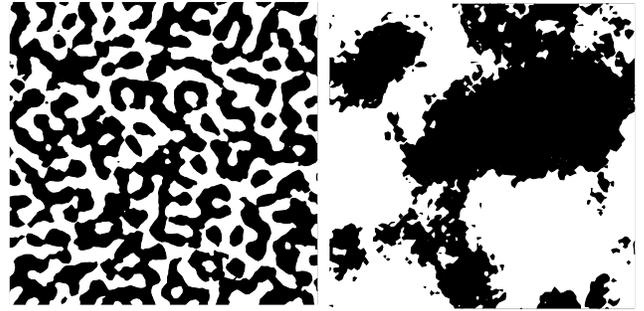}
\end{center}
\caption{Visualization of a 2D slice of size $150a$ of the oil-water density
configurations corresponding to regions C (left)
and B (right) in Fig. 1. The black (white) regions correspond to polar (hydrophobic) species}
\label{fig3} \end{figure}

A Hartree-Fock analysis of a model with energy $E=E_{0}+V$ yields at high
temperatures, $T\gtrsim T_{c}^{0}$, a homogeneous mixed phase (region A in
Fig.1). For lower temperatures $T\lesssim T_{c}^{0}$ (i.e. $t<1$) the
competition between $E_{0}$ and $V$ leads to essentially two distinct
regimes: either to a macroscopic phase separation in form of an emulsion
(region B in Fig.1) or to a fluid with micro-phase separation (region
C in Fig.1). For the latter to be stable within Hartree-Fock theory one
needs large surfactant molecules with not too small volume fractions at not
too low temperatures (at least $t\gtrsim \frac{a}{r_{{\rm s}}})$.
Consequently, micro-phase separation into a microemulsion occurs only for
intermediate temperatures $\frac{a}{r_{{\rm s}}}\lesssim t\lesssim 1$, a
behavior caused by the fact that the frustrating interaction, Eq.\ref{ham11}%
, is entropy-driven and vanishes as $T\rightarrow 0$. The typical size of an
oil or water micro-phase is $l_{m}$; these regions of given size are then
correlated over a distance $\xi $. For a sufficiently strong correlated
fluid, $\xi >l_{m}$, the density-density correlation function $G({\bf q}%
)=T^{-1}\left\langle \rho _{{\bf q}}\rho _{-{\bf q}}\right\rangle $ is given by
\begin{equation}
G(q)\ =\frac{q^{2}+q_{{\rm D}}^{2}\ }{\left( q^{2}-q_{m}^{2}\right)
^{2}+\left( 2 q_{m}/\xi\right)^{2}},  \label{Gapprox}
\end{equation}
with $q_{m}=\frac{2\pi }{l_{m}}$. In the limit of large $\xi$, the Fourier transformation
gives $G\left( x\right) =\frac{q_{m}^{2}+q_{{\rm D}}^{2}}{8\pi q_{m}x/\xi }%
e^{-x/\xi }\sin (\frac{2\pi x}{l_{m}})$, which clearly demonstrates our
physical interpretation of $\xi $ and $l_{m}$. Within Hartree-Fock theory
$\xi $ and $l_{m}$ are determined by the equation $G(q=0)^{-1}=\frac{6}{a^{2}}\left(
fta/r_{s}-1\right) +uTG\left( x=0\right) $. An illustrative visualization of
the various regimes can be obtained by using the clipped random wave
analysis of Ref.\cite{Choy00}. In Fig. 2 we show two typical
configurations, for the regions B and C of Fig.1, obtained through this analysis.

Based on these results for the equilibrium's behavior we will now study
glass formation in amphiphilic systems. A glass transition was recently
identified in a model for modulated charge inhomogeneities (``stripes'') in
doped Mott insulators, which is very similar to Eq.\ref{ham00},\ref{ham11},
however with infinite screening length. \cite{SW00,WSW01} It was shown that
self generated vitrification occurs if the charge correlation length, $\xi $%
, exceeds a critical value. These results were obtained for an infinite
range potential. In what follows we will apply the theoretical framework of
Ref.\cite{SW00,WSW01} to investigate vitrification for the model, Eq.\ref
{ham00},\ref{ham11}, with finite range interactions. We start from a
Cahn-Hilliard equation\cite{nigel92}
\begin{equation}
\gamma \frac{\partial \rho \left( {\bf x},t\right) }{\partial t}=\nabla^2\left(\frac{%
\delta E\left[ \rho \right] }{\delta \rho \left( {\bf x},t\right) }\right)+\eta
\left( {\bf x},t\right)  \label{Langevin}
\end{equation}
with damping coefficient, $\gamma $, and random $\eta \left( {\bf %
x},t\right) $, with white-noise correlation $\left\langle \eta \left( {\bf x}%
,t\right) \eta \left( {\bf x}^{\prime },t^{\prime }\right) \right\rangle
=-2\gamma T\nabla^2\delta \left( {\bf x-x}^{\prime }\right) \delta \left( t-t^{\prime
}\right) $ which simulates the effects of a thermal bath at temperature $T$.
We then determine the dynamical equations
for the retarded response function, as well as the correlation function of
the system within the self consistent screening approximation \cite{WSW01,Bray74}.
The onset of glassiness is signaled by a finite Edwards-Anderson order parameter
\begin{equation}
F\left( {\bf q}\right) =T^{-1}\lim_{t^{\prime }\rightarrow \infty
}\lim_{t\rightarrow \infty }\left\langle \rho \left( {\bf q},t+t^{\prime
}\right) \rho \left( {\bf -q},t^{\prime }\right) \right\rangle \ .
\label{EA}
\end{equation}
Note that the correlation function, Eq.\ref{EA}, vanishes in the equilibrium
fluid state. The solution of Eq.\ref{Langevin} using the weak ergodicity
breaking approach of Ref.\cite{CK93} (neglecting time derivatives in the
long time limit \cite{Claudio01}) yields a nonlinear self consistent
equation for $F\left( {\bf q}\right)$. The same result is obtained by
determining the configurational entropy via the replica approach of Refs.
\cite{Mon95,MP991}, where the glass transition is signaled by a
proliferation of metastable states, which become exponentially many with the
system size. We then find the region in the parameter space with finite Edwards
Anderson order parameter, $F\left( {\bf q}\right) $. We obtain that glassiness occurs
if the system undergoes micro-phase separation into a microemulsion (i.e. is located in the region
C of Fig.1) and if it forms a strongly correlated fluid of micro-phases with $\xi > 2l_{m}$. This rather simple criterion for glassiness, obtained by
analyzing the dynamical correlations and the spectrum of metastable states, along the lines of Refs. \cite{SW00,WSW01},
is the central result of this paper.

The boundary of the glassy state can
then be obtained by determining the dependence of $\xi $ and $l_{m}$ on $t,f,
$ and $r_{{\rm s}}/a$. Since the glass occurs for a fairly moderate
correlation length, we can ignore effects due to critical fluctuations
(relevant as $\xi \rightarrow \infty $) and use our results for $\xi $ and $%
l_{m}$ of the Hartree Fock analysis to determine the boundary of the
microemulsion glass state. This leads to the following implicit equation for
the onset temperature of glassiness $T_{A}=t_{A}T_{c}^{0}$:
\begin{equation}
f=\frac{\left( 1+B\left( t_{A},\frac{r_{{\rm s}}}{a}\right) f\right) ^{2}}{%
\frac{r_{{\rm s}}}{a}t_{A}}
\end{equation}
where $B\left( t,x\right) =\frac{6^{1/3}\pi ^{8/3}t^{3}}{x\left( t-1\right)
^{2}}$. The solution of this equation for $r_{{\rm s}}=10a$ gives the
boundary of the microemulsion glass shown in Fig.1; a more detailed analysis
for various ratios $r_{{\rm s}}/a$ is shown in Fig.3, where we also present
the dependence of the point $f^{\ast }$
for the onset of glassiness as function of $r_{{\rm s}}$. Since the
frustrating potential, Eq.\ref{ham11}, is of finite range, a finite critical
strength of the frustration, $\sim f^{\ast }$, must be reached to form a
microemulsion glass. Only as $r_{{\rm s}}\rightarrow \infty $ is there a glass
for arbitrarily small frustration\cite{SW00,WSW01}. The larger the
amphiphilic molecule and the larger its volume fraction, the more pronounced
is the glass state. In these limits, it has also been shown\cite{Woo96} that
the bending rigidity of amphiphilic membranes grows considerably, supporting
our results for a mechanically stable (high viscosity) glass.

\begin{figure}[t]
\begin{center}
\leavevmode \epsfxsize=8cm \epsffile{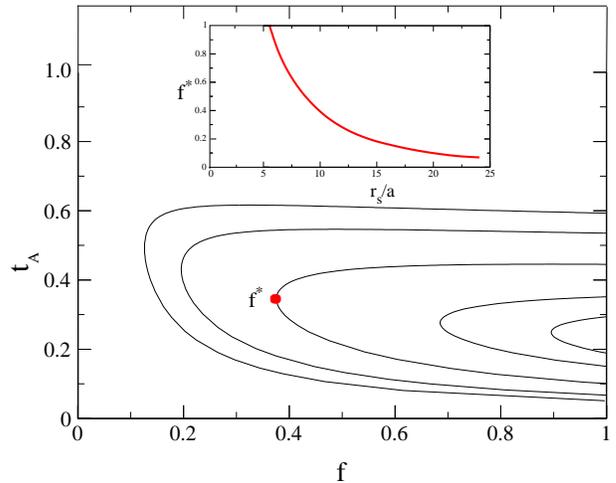}
\end{center}
\caption{Glass transition boundaries for different sizes, $r_s$ of amphiphilic
molecules ($r_s = 20,15,10,7,6$ from left to right). Inset: Temperature dependence of the critical volume fraction, $f^*$, for glassiness as a function of the size of the amphiphilic molecule} \label{fig2}
\end{figure}

The glassy state of our theory is self generated, i.e., it is due to the frustrated
nature of the interactions of the system. One way to test our theory is to deliberately
``lift'' this frustration. For example, we expect that if some disorder is imposed in the surfactant
chain length, the glass formation might be reduced. In fact, glassiness will disappear if we add
a certain amount of short chain amphiphiles to a system with long chain surfactants. Since the
effective mean square root length of the surfactant molecule is then given by $r_{{\rm s}}=\left(
\sum_{i}x_{i}r_{{\rm s,}i}^{2}\right) ^{1/2}$, with  $x_{i}$ being the mole fraction
of a surfactant with length $r_{{\rm s},i}$, $r_{{\rm s}}$ can easily become
smaller than the critical value $r_{{\rm s}}^{\ast }$. Another way of testing our
theory is to compare the transitions on the high-$T$ and low-$T$ side of the
glass state for given $f$. The low-$T$ transition has a larger modulation
length. Thus, it has a smaller configurational entropy ($S_{c}\propto
l_{m}^{-3}$),\cite{WSW01} compared to the high-$T$ side of the transition,
an effect which can be observed by measuring the specific heat anomaly at the
vitrification. At the same time, along the glass transition curve $\xi /l_{m}\simeq 2$.
Thus, the position of a small angle neutron scattering
(SANS) peak should be shifted along the transition curve, whereas the peak-width ratio should stay
essentially unchanged.

In a recent experiment\cite{perreur01} the rheological and structural (as
probed by SANS) properties of an aqueous solution of a four branched
copolymer were studied and a phase diagram similar to our Fig.1, with a
``stiff gel'' at the position of our microemulsion glass was found.
It was shown that above a critical volume fraction
of the polymer ($f\simeq 0.3$), the viscosity of the emulsion diverges at a
finite temperature which is a monotonically decreasing function of $f$.
Furthermore, the divergence of the viscosity is associated with an enhancement
of the liquid correlations in a form consistent with the criteria for
glassiness proposed in this paper ($\xi \approx 2l_{m}$). All these
experimental findings seem to be in agreement with the theory proposed here,
even though this system is actually a mixture of block copolymers and
water only, without the oil. We argue that the ``quasi-solid''
or ``stiff-gel''\cite{glatter94} found in Ref.\cite{perreur01} is in fact a self generated glass
formed by correlated structures of the microemulsion.

In conclusion, we have shown that the competition between entropic effects
and stoichiometric constraints responsible for the formation of micelles in
microemulsions can lead also to the emergence of a self generated glassy
behavior in these systems. This effect is prominent when the effective size
of the amphiphile molecules, $r_{{\rm s}}$, exceeds a critical value of $6a$%
, where $a$ is the typical correlation length of water and oil. Furthermore,
we show that there is a critical volume fraction to achieve the glassy
behavior which depends solely on $\frac{r_{{\rm s}}}{a}$. Owing to the smaller
energy densities at the larger length scales relevant to
them, nonlinearity is easier to achieve in the laboratory for microemulsion
glasses than the usual structural glasses. In fact, the mechanical
properties of the microemulsion glass proposed here are likely much closer to
soft materials\cite{cloitre00,sollich97,mason95} such as gels and pastes
than to a rigid window glass. Nevertheless, the universality class of the
microemulsion glasses is identical to the one which is believed to apply to
structural glasses\cite{KTW89}, a conclusion which is also supported by the
results of Refs.\cite{cloitre00,sollich97}. The fact that the parameters
governing the emergence of glassiness in these systems are easily tunable,
makes them perfect probes for the entropy crisis scenario attributed to
glasses.

We are grateful to D. Oxtoby for useful discussions. This research was
supported by the Institute for Complex Adaptive Matter, an award from
Research Corporation (J.S.), the Ames Laboratory, operated for the U.S.
Department of Energy by Iowa State University under Contract No.
W-7405-Eng-82 (S. W., H.W.Jr. and J. S.), and the National Science
Foundation grant CHE-9530680 (P. G. W.). H.W.Jr. also acknowledges support
from FAPESP.

\end{document}